\documentclass{sbrt}
\usepackage[english, portuguese]{babel}
\usepackage[utf8]{inputenc}

\usepackage{amsmath,amssymb,amsfonts}
\usepackage{graphics,graphicx}
\usepackage{epsfig}
\usepackage{cite}
\usepackage{bm}
\usepackage{times}
\usepackage{subcaption}
\usepackage[T1]{fontenc}
\usepackage{subcaption}
\usepackage{color}
\usepackage{algorithm}
\usepackage{algpseudocode}
\usepackage{algorithmicx}
\usepackage{enumerate}
\usepackage{epstopdf}
\usepackage{lipsum} 
\usepackage{stmaryrd}
\usepackage{algpseudocode}
\usepackage{url}
\usepackage{balance}
\usepackage{textcomp}
\usepackage{stfloats}
\usepackage{array}
\usepackage{multirow}
\usepackage{rotating}
\newtheorem{proposition}{Proposition}
\newtheorem{remark}{Remark}
\newtheorem{demonstration}{Proof}
\newcommand{\Mod}[1]{\ (\mathrm{mod}\ #1)}
\begin{document}
\bstctlcite{BSTcontrol}

% \title{\textit{Dimming Space-Time Code} (DSTC) para Comunicação por Luz Visível (VLC) com Estimação Semi-Cega de Símbolos e Canal}
% \title{\textit{Dimming Space-Time Code} (DSTC): Comunicação por Luz Visível com Estimação Semi-Cega de Símbolos e Canal}
\title{Dimming Space-Time Code (DSTC) for Visible Light Communication with Semi-Blind Detection}

\author{Igor S. C. Rodrigues, Leandro R. Ximenes e André L. F. de Almeida}

\maketitle

\begin{abstract}
Visible light communication (VLC) provides a unified framework for wireless data transmission and illumination, but its practical deployment requires transmission schemes that jointly satisfy communication and lighting constraints. In color-shift keying (CSK) systems, dimming remains a challenging and underexplored problem because the average optical power must be controlled without altering the perceived chromaticity. This paper proposes a dimming space-time code (DSTC) for CSK-based VLC systems, in which a structured dimming matrix introduces controlled temporal power variations while satisfying physical feasibility, color-preservation, and identifiability conditions. Two receiver architectures are developed: a pilot-assisted zero-forcing (ZF) receiver and a tensor-based semi-blind PARAFAC receiver that jointly estimates the channel and transmitted symbols using only one training time slot. Simulation results show that the proposed DSTC provides diversity gains and substantial BER reductions with respect to conventional CSK, while the tensor-based receiver improves spectral efficiency by reducing training overhead, with particular benefits in large-scale MIMO configurations.
\end{abstract}

\begin{keywords}
Visible Light Communication, color-shift keying, dimming control, tensor modeling.
\end{keywords}

\renewcommand\baselinestretch{.9}

\section{Introduction}
Visible light communication (VLC) has emerged as a promising technology for simultaneous wireless data transmission and illumination by exploiting the unlicensed visible-light spectrum. In comparison with radio-frequency (RF) systems, VLC can provide high data rates, low-cost implementation, spatial confinement, and enhanced physical-layer security \cite{OBrien2008}. Since VLC transmitters are also illumination sources, communication design must account for lighting constraints in addition to conventional reliability and spectral-efficiency requirements. In particular, dimming control, i.e., the adjustment of the average optical power emitted by the LEDs, is required to satisfy user comfort and energy-efficiency targets while avoiding perceptible flicker and performance degradation \cite{Rajagopal2012}. Most existing dimming techniques have been developed for intensity-modulation/direct-detection (IM/DD) schemes such as on-off keying (OOK) and pulse-position modulation (PPM), where dimming is typically implemented through duty-cycle adaptation or symbol-probability shaping \cite{Lee2011, Guo2022, Babalola2022, Babalola2024}.

Color-shift keying (CSK) conveys information by controlling the relative optical intensities of LEDs with distinct chromatic components, so that data symbols are represented by points in a chromaticity diagram \cite{Rajagopal2012}. This modulation principle is attractive for VLC because it enables illumination with reduced flicker and improved energy efficiency. However, dimming in CSK systems is more constrained than in scalar IM/DD schemes: the average optical power must be adjusted without perturbing the target chromaticity perceived by the user. Consequently, dimming control for CSK remains relatively underexplored, especially in multiple-input multiple-output (MIMO) configurations. The multidimensional design of CSK-based MIMO-VLC systems, potentially involving spatial, temporal, and chromatic dimensions, suggests that dimming can also be incorporated into the transmission code as an additional degree of freedom. In this perspective, controlled power variations across LEDs and time slots can provide diversity while preserving the average emitted color. Existing CSK diversity schemes, such as the color-hopping space-time (CHST) code in \cite{9612407}, exploit color, spatial, and temporal diversity, but they do not explicitly formulate dimming as a coding dimension nor address the corresponding receiver structure.

Tensor modeling provides a rigorous algebraic framework for communication systems whose signals are naturally indexed by multiple coupled dimensions. Parallel factor analysis (PARAFAC) and related tensor decompositions have been widely used for blind and semi-blind receiver design, multiuser equalization, constrained MIMO factorization, space-time-frequency coding, and reconfigurable-surface-assisted systems, benefiting from identifiability properties that can reduce training overhead \cite{de2007parafac,de2008constrained,FAVIER2012,Favier_EUSIPICO,XIMENES2023104192,10220086,10639165,ximenes2015semi,deAraujoSAM2020,Araujo2021,ardah2021trice}. These properties are particularly relevant for CSK-based VLC with dimming codes, because the received signal can be represented through coupled spatial, temporal, and chromatic factors associated with the channel, the transmitted symbols, and the dimming pattern.

Motivated by these observations, this paper proposes a dimming space-time code (DSTC) for CSK-based VLC systems. The proposed code introduces controlled temporal variations in optical power through a dimming matrix that satisfies physical nonnegativity, average-power, chromaticity-preservation, and rank conditions. Unlike conventional dimming approaches, the proposed DSTC uses dimming not only as an illumination-control mechanism but also as a source of transmission diversity. Moreover, two receiver strategies are developed: a pilot-based zero-forcing (ZF) receiver and a semi-blind tensor receiver based on PARAFAC/Khatri--Rao factorization, which jointly estimates the channel and transmitted symbols with only one training time slot. To the best of the authors' knowledge, this is the first work to exploit dimming as a diversity dimension in CSK-based VLC and to introduce a tensor-based semi-blind receiver for this class of systems.

The remainder of this paper is organized as follows. Section 2 presents the proposed DSTC signal model and dimming-pattern design. Section 3 derives the ZF and tensor-based receivers. Section 4 discusses the simulation results, including bit-error-rate, channel-estimation, and spectral-efficiency comparisons. Finally, Section 5 concludes the paper.

\section{Dimming Space-Time Code (DSTC)}

\begin{figure}[t!]
	\centering
	\includegraphics[width=0.5\textwidth]{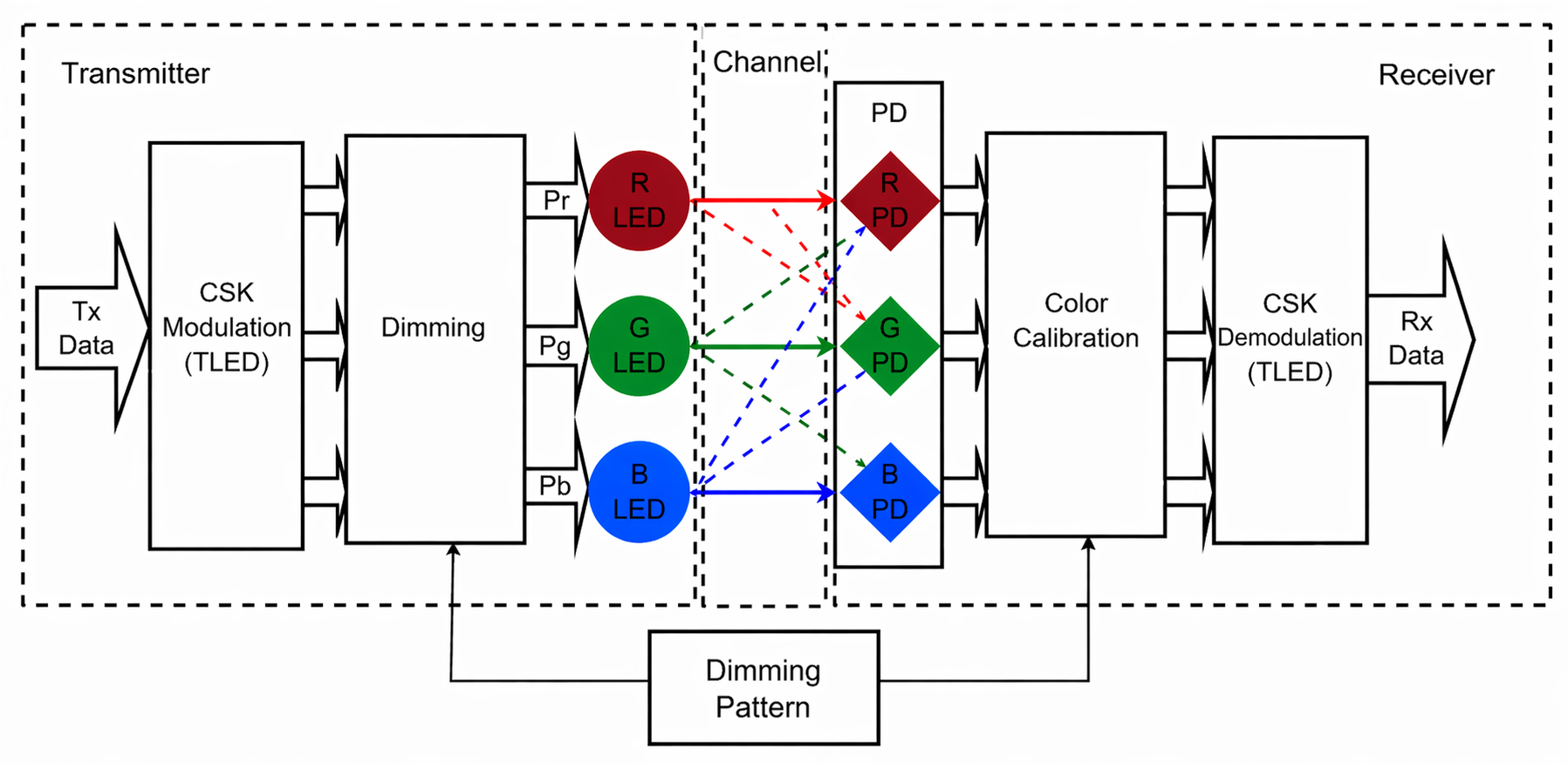}
	\caption{Proposed DSTC transmission scheme for a SISO TLED/CSK VLC system.}
	\label{fig:tx_diagram}
\end{figure}

Consider a point-to-point multiple-input multiple-output (MIMO) VLC block-transmission system employing CSK modulation under a linear intensity-modulation/direct-detection (IM/DD) channel model. The transmitter consists of $L_T$ LED groups, each containing $K_T$ color channels, yielding a total of $ K_TL_T$ transmit optical sources. The symbol matrix $\mathbf{S} \in \mathbb{R}^{N \times K_TL_T}$ contains a block of $N$ CSK symbols multiplexed across these optical sources. For instance, in the QuadLED (QLED) architecture proposed in \cite{Singh2014}, $K_T = 4$ color channels are considered.
The goal of the proposed DSTC scheme is to impose controlled temporal variations on the instantaneous optical power while preserving the prescribed average color and illumination level. This is accomplished through a dimming matrix $\mathbf{C}$, whose entries scale the optical intensity emitted by each LED at each transmission state. As illustrated in Fig.~\ref{fig:tx_diagram}, the dimming block therefore introduces an additional temporal coding dimension that can be exploited for diversity without changing the average chromaticity perceived by the receiver.

Assume that the block of $N$ symbols in $\mathbf{S}$ is transmitted over $K$ dimming states. The dimming matrix $\mathbf{C} \in \mathbb{R}^{K \times K_TL_T}$ is defined such that its $k$-th row specifies the normalized dimming coefficients applied to all transmit LEDs during the $k$-th state. Let $\mathbf{\Psi}_k(\mathbf{C}) \in \mathbb{R}^{K_TL_T \times K_TL_T}$ denote the diagonal matrix obtained from the $k$-th row of $\mathbf{C}$, i.e.,
\begin{align}
\mathbf{\Psi}_k
=
\operatorname{diag}(C_{k,1}, C_{k,2}, \dots, C_{k,K_TL_T}).
\label{eq:dimming_diag_definition}
\end{align}
The transmitted signal matrix at the $k$-th dimming state is
\begin{align}
    \mathbf{X}_k = \mathbf{\Psi}_k\mathbf{S}^T,
    \label{eq:XS}
\end{align}
where $\mathbf{X}_k \in \mathbb{R}^{K_TL_T \times N}$. Thus, the coefficients in $\mathbf{\Psi}_k$ scale the instantaneous optical power of each LED, whereas the design of $\mathbf{C}$ enforces the desired average illumination and chromaticity over the complete set of $K$ states.

Fig.~\ref{fig:tx_diagram} illustrates the DSTC principle for a single-input single-output (SISO) configuration ($L_R = L_T = 1$) using triLED (TLED) transmitters and receivers ($K_R = K_T = 3$).

\subsection{Dimming Pattern Design}
The dimming matrix $\mathbf{C}$ must satisfy two sets of requirements: physical feasibility constraints and an algebraic rank condition.

\textit{1) Physical constraints:} Since the entries of $\mathbf{C}$ represent normalized optical-intensity scaling factors under an IM/DD model, they must satisfy
\begin{align}
0 \le C_{k,i} \le 1, \quad \forall k,i,
\label{eq:nnC}
\end{align}
where $i = 1,\dots,K_TL_T$ indexes the transmit LEDs. In addition, to impose a prescribed average optical power per LED, the coefficients must satisfy
\begin{align}
\frac{1}{K}\sum_{k=1}^{K} C_{k,i} = P_m, \quad \forall i,
\label{eq:pC}
\end{align}
where $P_m$ denotes the target normalized dimming level. The perceived color of illumination must also remain invariant on average. The resulting chromaticity coordinates are given by
\begin{align}
x &= \sum_{k_t=1}^{K_T} P_{k_t} x_{k_t}, 
&y = \sum_{k_t=1}^{K_T} P_{k_t} y_{k_t},
\label{eq:xyC}
\end{align}
where $\left(x_{k_t},y_{k_t}\right)$ are the CIE 1931 chromaticity coordinates of the $k_t$-th color channel and $P_{k_t}$ denotes its average transmitted optical power after dimming. Hence, $\mathbf{C}$ must be designed so that dimming changes the average optical power while preserving the target chromaticity.

textit{2) Rank condition:} From an algebraic viewpoint, $\mathbf{C}$ must have full column rank, i.e.,
\begin{align}
\operatorname{rank}(\mathbf{C}) = K_T L_T.
\label{eq:rC}
\end{align}
This condition requires $K \ge K_TL_T$ and ensures that the temporal dimming signatures assigned to different LEDs are linearly independent. It also guarantees the existence of a left pseudoinverse of $\mathbf{C}$, which is required by the proposed receiver architectures to separate the contributions from the different transmit optical sources.

\subsection{Construction of $\mathbf{C}$ via Hadamard Matrix}

A structured construction satisfying the above requirements can be obtained from a Hadamard matrix. Specifically, define
\begin{align}
\mathbf{C} = P_m \mathbf{1}_{K\times K_TL_T} + \alpha \mathbf{B},
\label{eq:Chadamard}
\end{align}
where $\mathbf{1}_{K\times K_TL_T}$ is an all-ones matrix and $\mathbf{B} \in \mathbb{R}^{K \times K_TL_T}$ is formed by selecting columns from a Hadamard matrix of order $K$, with entries in $\{-1,+1\}$. The scalar $\alpha$ controls the amplitude of the temporal power variations around the average dimming level $P_m$.

One possible construction is to select $K_TL_T$ nonconstant columns from a Hadamard matrix of order $K$, excluding the all-ones column. These selected columns have zero sample mean, i.e., $\sum_{k=1}^{K} B_{k,i} = 0$, which directly enforces \eqref{eq:pC}. Since at most $K-1$ nonconstant Hadamard columns are available, this construction requires $K_TL_T \le K-1$. In addition, standard Hadamard matrices exist for orders satisfying $K=1$, $K=2$, or $K \Mod 4 = 0$; in the present design, $K$ is chosen so that a suitable order-$K$ Hadamard matrix is available.
To ensure nonnegative and bounded optical intensities, $\alpha$ must satisfy
\begin{align}
\alpha \le \min(P_m,\; 1 - P_m),
\end{align}
which ensures \eqref{eq:nnC} since each entry of $\mathbf{C}$ lies in $\{P_m-\alpha, P_m+\alpha\}$. The zero-mean property of these Hadamard columns guarantees \eqref{eq:pC}, whereas their mutual orthogonality ensures \eqref{eq:rC}.

\section{Symbol Detection}

This section presents the proposed receiver architectures for symbol detection in the DSTC scheme. First, we describe a conventional \textit{zero-forcing} (ZF) receiver based on pilot-assisted channel estimation. We then introduce a tensor-based semi-blind receiver that exploits the PARAFAC structure of the received signal, enabling joint channel and symbol estimation with reduced training overhead. At the receiver, $L_R$ groups of $K_R$ photodetectors are employed. The received signal associated with the $k$-th dimming state is obtained by transmitting the signal in \eqref{eq:XS} through the wireless optical channel, yielding
\begin{align}
\mathbf{Y}_k = \mathbf{H}_k\mathbf{X}_k
&=
\mathbf{H}_k\mathbf{\Psi}_{k}\mathbf{S}^T + \mathbf{N}_k,
\label{eq:received_matrix_dimming_state}
\end{align}
where $\mathbf{H}_k \in \mathbb{R}^{K_RL_R \times K_TL_T}$ denotes the channel matrix, $\mathbf{N}_k$ is the additive white Gaussian noise (AWGN) matrix, and $\mathbf{Y}_k \in \mathbb{R}^{K_RL_R \times N}$ denotes the received symbol matrix.

\subsection{Zero-Forcing Receiver}

For the ZF receiver, the $K$ received blocks in \eqref{eq:received_matrix_dimming_state} are stacked into the concatenated matrix $\mathbf{\bar{Y}} \in \mathbb{R}^{K_RL_RK \times N}$, defined as
\begin{align}
\mathbf{\bar{Y}} 
&= \left[\begin{array}{c}
\left(\mathbf{H}_{1}\mathbf{\Psi}_{1}\right)^T
\cdots
\left(\mathbf{H}_{K}\mathbf{\Psi}_{K}\right)^T
\end{array}\right]^T\mathbf{S}^T + \mathbf{N}.
\label{H_eff*S}
\end{align}
From this, the effective channel matrix $\mathbf{H}_e$ is defined as
\begin{align}
 \hspace{-2ex}\mathbf{H}_{e} = \textit{blkdiag}\left(\mathbf{H}_{1},\mathbf{H}_{2},\cdots,\mathbf{H}_{K}\right)\left[\mathbf{\Psi}_{1}^T \; \mathbf{\Psi}_{2}^T \; \cdots \;\mathbf{\Psi}_{K}^T\right]^T,
 \label{H_eff} 
\end{align}
where $\textit{blkdiag}(\cdot)$ denotes a block-diagonal matrix. Substituting \eqref{H_eff} into \eqref{H_eff*S}, the received signal model can be rewritten as
\begin{align}
    \mathbf{\bar{Y}} = \mathbf{H}_e\mathbf{S}^T + \mathbf{N}.
    \label{eq:Ymimo}
\end{align}

Channel estimation can be performed by transmitting a pilot sequence before data transmission. Assuming the channel remains constant across the $K$ dimming states, a simple strategy is to activate one LED at a time at maximum power while keeping the others off during each time slot. This procedure requires $K_TL_T$ pilot time slots, which reduces spectral efficiency and effective throughput.

The pilot symbol sequence is known a priori at both the transmitter and the receiver and is defined by the matrix $\mathbf{S}_{0}$. Using \eqref{eq:Ymimo}, the receiver estimates the channel matrix as
\begin{align}
\mathbf{\widehat{H}}_e \triangleq \mathbf{\tilde{Y}}_{0} \left(\mathbf{S}_{0}^{T}\right)^{\dagger},
\label{eq:Chest}
\end{align}
where $\mathbf{\tilde{Y}}_{0}$ is the set of received signals during the channel estimation phase, and $\dagger$ represents the Moore--Penrose pseudoinverse.
Since the dimming matrix $\mathbf{C}$ is also known at the receiver, an equalizer can be applied to estimate $\mathbf{S}$ from \eqref{eq:Ymimo} and \eqref{eq:Chest}. The resulting ZF equalizer is given by
\begin{align}
    \mathbf{\hat{S}} = \left(\mathbf{\widehat{H}}_{e}^{\dagger}\mathbf{\bar{Y}}\right)^T.
    \label{S_ZF}
\end{align}
%where $\mathbf{\hat{S}}$ is the estimated $\mathbf{S}$, and $\mathbf{\widehat{H}}_{e}$ is the estimated effective channel matrix.

\subsection{Semi-Blind Tensor-Based Channel and Symbol Estimator}

Alternatively, the received data can be rearranged into a third-order tensor model, enabling semi-blind joint channel and symbol estimation through the PARAFAC decomposition originally introduced in \cite{HARSHMAN199439}. Using the identity
\begin{align}
\operatorname{vec}(\mathbf{A}\mathbf{X}\mathbf{B})
=
(\mathbf{B}^T \otimes \mathbf{A})\operatorname{vec}(\mathbf{X}),
\end{align}
and assuming that the channel remains constant over $K$ dimming states, the vectorized received signal is given by
\begin{align}
\mathbf{y}_k
=
(\mathbf{S} \otimes \mathbf{H})
\operatorname{vec}(\mathbf{\Psi}_k)
\in \mathbb{R}^{K_RL_RN \times 1}.
\label{eq:vec_received_state}
\end{align}
Since $\mathbf{\Psi}_k$ is diagonal, it follows that
\begin{align}
(\mathbf{S} \otimes \mathbf{H})
\operatorname{vec}(\mathbf{\Psi}_k)
=
(\mathbf{S} \diamond \mathbf{H})\mathbf{c}_k^T,
\label{eq:khatri_rao_identity_dimming}
\end{align}
where $\mathbf{c}_k$ denotes the $k$-th row of $\mathbf{C}$. Thus,
\begin{align}
\mathbf{y}_k
=
(\mathbf{S} \diamond \mathbf{H})\mathbf{c}_k^T.
\label{eq:vec_received_final}
\end{align}
Finally, stacking all dimming states yields
\begin{align}
\mathbf{Y}_{(3)} =
\begin{bmatrix}
\mathbf{y}_1^T 
\mathbf{y}_2^T
\cdots
\mathbf{y}_K^T
\end{bmatrix}^T
=
\mathbf{C}(\mathbf{S} \diamond \mathbf{H})^T
\in \mathbb{R}^{K \times K_RL_RN}.
\label{eq:stacked_vectorized_model}
\end{align}
This structure naturally leads to the PARAFAC decomposition
\begin{align}
\mathcal{Y} = [\![\mathbf{H},\mathbf{S},\mathbf{C}]\!] \in \mathbb{R}^{K_RL_R \times N \times K}.
\label{ytensor_CP}
\end{align}

The three modes correspond, respectively, to receiver elements, transmitted symbols, and dimming states. The PARAFAC structure yields the following unfoldings of $\mathcal{Y}$:
\begin{align}
\mathbf{Y}_{(1)}
&=
\mathbf{H}
(\mathbf{C} \diamond \mathbf{S})^T
\in \mathbb{R}^{K_RL_R \times NK},
\label{eq:mode1_unfolding}\\
\mathbf{Y}_{(2)}
&=
\mathbf{S}
(\mathbf{C} \diamond \mathbf{H})^T
\in \mathbb{R}^{N \times K_RL_RK},
\label{eq:mode2_unfolding}\\
\mathbf{Y}_{(3)}
&=
\mathbf{C}
(\mathbf{S} \diamond \mathbf{H})^T
\in \mathbb{R}^{K \times K_RL_RN}.
\label{eq:mode3_unfolding}
\end{align}

\subsubsection{Uniqueness Conditions}
\label{prop:kruskal_condition_S2CCHST}

For the tensor model to be identifiable, and hence enable semi-blind joint channel and symbol estimation, the PARAFAC decomposition must be essentially unique up to permutation and scaling.

\begin{proposition}
\label{prop:kruskal_video}
Assume that $\mathbf{S}$ has full column rank and that $\mathbf{C}$ is constructed according to \eqref{eq:Chadamard}. Then, (\ref{ytensor_CP}) is essentially unique if $k_H \geq 2$, where $k_H$ is the Kruskal rank of $\mathbf{H}$.
\end{proposition}

\begin{demonstration}
According to Kruskal’s condition \cite{kruskal1977three}, for a model of rank $R = K_T L_T$, it holds that $k_H + k_S + k_C \geq 2R + 2$.
% \begin{equation}
% k_H + k_S + k_C \geq 2R + 2.
% \end{equation}
Since $\mathbf{S}$ has full column rank for sufficiently large $N$, we have $k_S = K_TL_T$. Moreover, because $\mathbf{C}$ is constructed from distinct columns of a Hadamard matrix as in \eqref{eq:Chadamard}, it also has full column rank, yielding $k_C = K_TL_T$. Therefore, the condition reduces to $k_H \geq 2$, which ensures essential uniqueness.
\end{demonstration}

\begin{remark}
While the ZF receiver requires $K_RL_R \geq K_TL_T$, the tensor-based approach only requires that no two rows of $\mathbf{H}$ be linearly dependent. Under the commonly adopted VLC assumption of negligible optical crosstalk, $\mathbf{H}$ is approximately diagonal, which naturally satisfies this requirement even in scenarios with fewer receivers than transmitters.
\end{remark}

\begin{proposition}
If $N < K_TL_T$ and $K_RL_R > K_TL_T$, then $k_S \geq 2$ is sufficient for essential uniqueness.
\end{proposition}

\begin{demonstration}
$\mathbf{H}$ has full column rank since it is a diagonal matrix with more rows than columns, yielding $k_H = K_TL_T$. Similarly, $\mathbf{C}$ has full column rank by construction, resulting in $k_C = K_TL_T$. Applying Kruskal’s condition leads to $k_S \geq 2$.
\end{demonstration}

\subsubsection{Khatri-Rao Factorization (VLC-KRF)}

Once the PARAFAC structure is shown to be identifiable, the channel and symbol matrices can be jointly estimated through Khatri-Rao factorization. From the mode-3 unfolding in \eqref{eq:mode3_unfolding}, we obtain
\begin{equation}
\mathbf{Q} = \left(\mathbf{Y}_{(3)}\right)^T\left( \mathbf{C}^T \right)^{\dagger} = (\mathbf{S} \diamond \mathbf{H}).
\label{What}
\end{equation}

The objective is to factorize each column $\widehat{\mathbf{q}}_r$, with $r = 1, \dots, K_T L_T$, as the Khatri-Rao product of two vectors $\widehat{\mathbf{s}}_r$ and $\widehat{\mathbf{h}}_r$, such that $\widehat{\mathbf{q}}_r \approx \widehat{\mathbf{s}}_r \otimes \widehat{\mathbf{h}}_r = \operatorname{vec}(\widehat{\mathbf{h}}_r\widehat{\mathbf{s}}_r^T)$.
% \begin{equation}
% \widehat{\mathbf{q}}_r \approx \widehat{\mathbf{s}}_r \otimes \widehat{\mathbf{h}}_r = \operatorname{vec}(\widehat{\mathbf{h}}_r\widehat{\mathbf{s}}_r^T).
% \end{equation}
Let $\mathbf{Q}_r$ denote the reshaped version of $\widehat{\mathbf{q}}_r$, defined as $\mathbf{Q}_r = \mathrm{unvec}_{K_RL_R \times N}(\widehat{\mathbf{q}}_r)
\approx \mathbf{h}_r \mathbf{s}_r^T $.
% \begin{equation}
% \mathbf{Q}_r = \mathrm{unvec}_{K_RL_R \times N}(\widehat{\mathbf{q}}_r)
% \approx \mathbf{h}_r \mathbf{s}_r^T .
% \end{equation}
Therefore, recovering $\mathbf{h}_r$ and $\mathbf{s}_r$ amounts to solving a rank-one approximation problem:
\begin{equation}
\min_{\mathbf{h}_r,\mathbf{s}_r} 
\left\|
\mathbf{Q}_r - \mathbf{h}_r \mathbf{s}_r^T
\right\|_F^2,
\end{equation}
which can be solved using the SVD. Algorithm~\ref{alg:VLC-KRF} summarizes the proposed VLC-KRF receiver. For an arbitrary $n$, the scaling ambiguity is resolved using the diagonal matrix $\mathbf{\Delta}$ given by $\mathbf{\Delta} \approx \left(\mathbf{S}_n \right) ./ \left(\widehat{\mathbf{S}}_n \right), \,
\widehat{\mathbf{S}} \leftarrow \widehat{\mathbf{S}} \mathbf{\Delta}, \quad
\widehat{\mathbf{H}} \leftarrow \widehat{\mathbf{H}} \mathbf{\Delta}^{-1},$
% \begin{align}
% \mathbf{\Delta} \approx \left(\mathbf{S}_n \right) ./ \left(\widehat{\mathbf{S}}_n \right), \quad
% \widehat{\mathbf{S}} \leftarrow \widehat{\mathbf{S}} \mathbf{\Delta}, \quad
% \widehat{\mathbf{H}} \leftarrow \widehat{\mathbf{H}} \mathbf{\Delta}^{-1},
% \label{eq:ambiguity_matrix}
% \end{align}
where $\mathbf{S}_n$ is a known row of the transmitted symbol matrix. Therefore, the tensor model requires only a single known row of $\mathbf{S}$, reducing the training sequence from $K_TL_T$ time slots in the ZF receiver to a single time slot.

\begin{algorithm}[!t]
\caption{VLC-KRF}
\begin{algorithmic}[1]
\Require $\Tilde{\mathcal{Y}}, \mathbf{C}, \mathbf{S}_n$
\State $\Tilde{\mathbf{Y}}^{(3)} \leftarrow$ mode-3 unfolding of $\Tilde{\mathcal{Y}}$
\State $\mathbf{Q} = \Tilde{\mathbf{Y}}^{(3)} (\mathbf{C}^T)^{\dagger} \approx \mathbf{S} \diamond \mathbf{H}$

\For{$r = 1, \ldots, K_TL_T$}
    \State $\mathbf{Q}_r = \mathrm{unvec}_{K_RL_R \times N}(\mathbf{q}_r)$
    \State $[\mathbf{U},\mathbf{\Sigma},\mathbf{V}] = \mathrm{svd}(\mathbf{Q}_r)$
    \State $\widehat{\mathbf{h}}_r = \mathbf{U}_{.1}, \quad \widehat{\mathbf{s}}_r = \mathbf{V}_{.1}$
\EndFor

\State $\mathbf{\Delta} = \mathbf{S}_n ./ \widehat{\mathbf{S}}_n$
\State $\widehat{\mathbf{S}} \leftarrow \widehat{\mathbf{S}}\mathbf{\Delta}, \quad
\widehat{\mathbf{H}} \leftarrow \widehat{\mathbf{H}}\mathbf{\Delta}^{-1}$

\Ensure $\widehat{\mathbf{H}}, \widehat{\mathbf{S}}$
\end{algorithmic}
\label{alg:VLC-KRF}
\end{algorithm}

\subsection{Spectral Efficiency} \label{subsec:SE}

For a $4$-CSK system, each symbol conveys $\log_2(4)=2$ bits. Considering $L_T$ transmitters and $N$ distinct symbols per block, the total number of useful transmitted bits is $2L_TN$. Since the $K$ transmission blocks correspond to repetitions of the same symbols, they only increase the transmission time.

For the ZF and VLC-KRF receivers, the spectral efficiencies are given by
\begin{align}
\eta_{\text{ZF}} &= \frac{2L_TN}{NK + K_TL_T}, \label{eq:SE1}\\
\eta_{\text{VLC-KRF}} &= \frac{2L_TN}{NK + 1}. \label{eq:SE2}
\end{align}
Thus, the spectral-efficiency gain of VLC-KRF is exclusively due to the reduced training overhead:
\begin{equation}
\frac{\eta_{\text{VLC-KRF}}}{\eta_{\text{ZF}}}
=
\frac{NK + K_TL_T}{NK + 1}. \label{eq:SE3}
\end{equation}
Therefore, the advantage of the tensor-based receiver is more pronounced for small values of $NK$, whereas both receivers achieve similar efficiencies as $NK$ increases.

\section{Results and Discussion}
\label{sec:simulations}
The channel $\mathbf{H}$ is modeled as a Gaussian random matrix that remains constant over the $K$ transmission blocks and is corrupted by AWGN. BER results are obtained through Monte Carlo simulations using $10^5$ equiprobable 4-CSK symbols transmitted in blocks of $N=100$ symbols. Unless otherwise stated, $P_m=0.5$ and $\alpha=0.4$.
Table \ref{tab:power_color} compares the average transmitted color and average transmitted power before and after applying the DSTC scheme. The average transmitted color is computed using \eqref{eq:xyC}. The results show that the proposed scheme achieves the desired dimming level without altering the average transmitted color, with variations remaining below the typical perceptual sensitivity predicted by Weber's law.

\begin{table}[!t]
\centering
\caption{Average power and color before and after dimming ($K_T{=}3$, $L_T{=}2$, $K{=}12$)}
\scriptsize
\setlength{\tabcolsep}{15pt}
\begin{tabular}{lcc}
\hline
Metric & Before & After \\
\hline
Power & 1 & 0.5001 \\
Color $(x,y)$ & $(0.4389,\,0.2910)$ & $(0.4389,\,0.2909)$ \\
\hline
\end{tabular}
\label{tab:power_color}
\end{table}

\begin{figure}[!t]
\centering
\subfloat[]{\includegraphics[width=0.8\columnwidth]{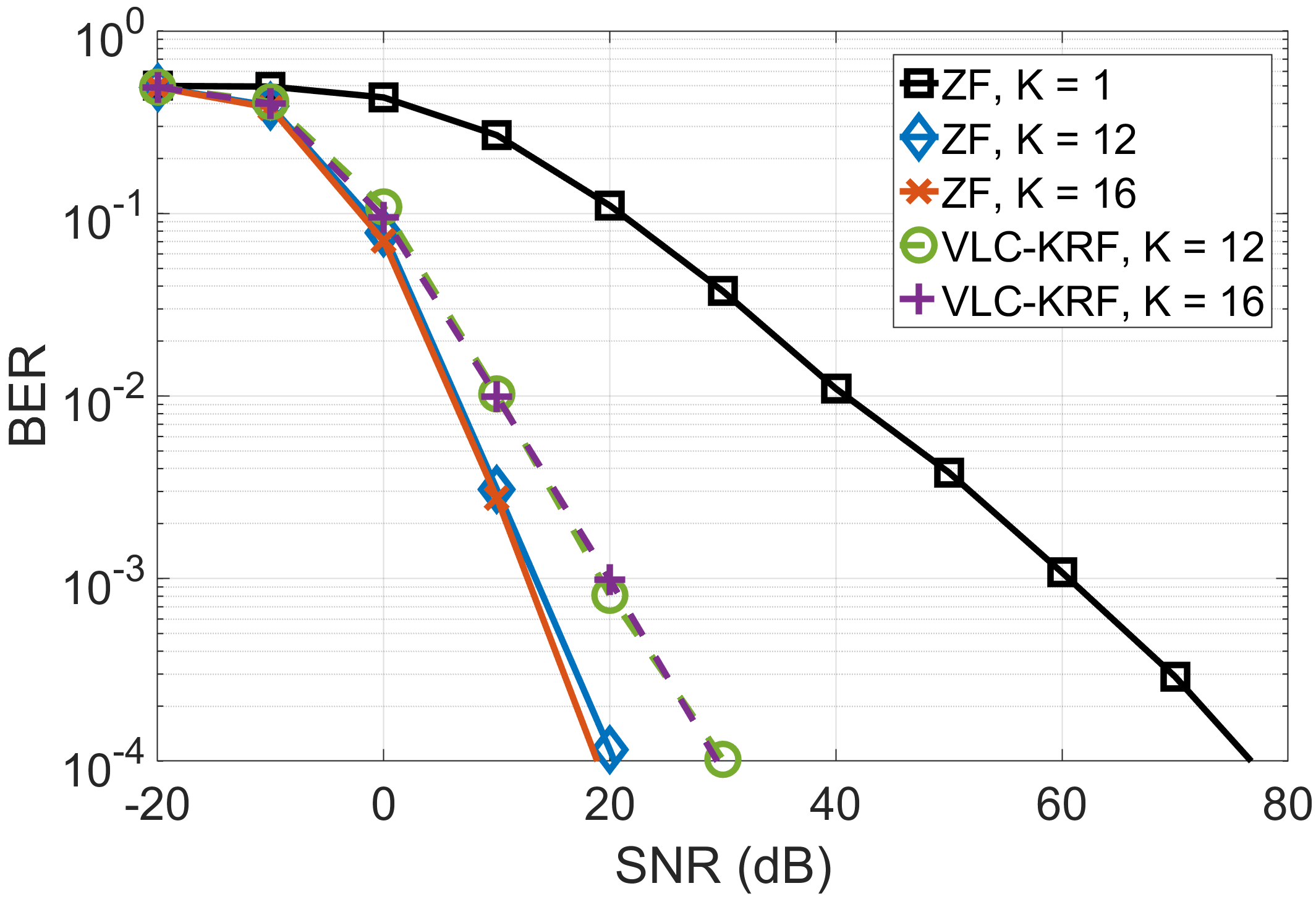}
\label{fig:SNRBER}}
\hfill
\subfloat[]{\includegraphics[width=0.8\columnwidth]{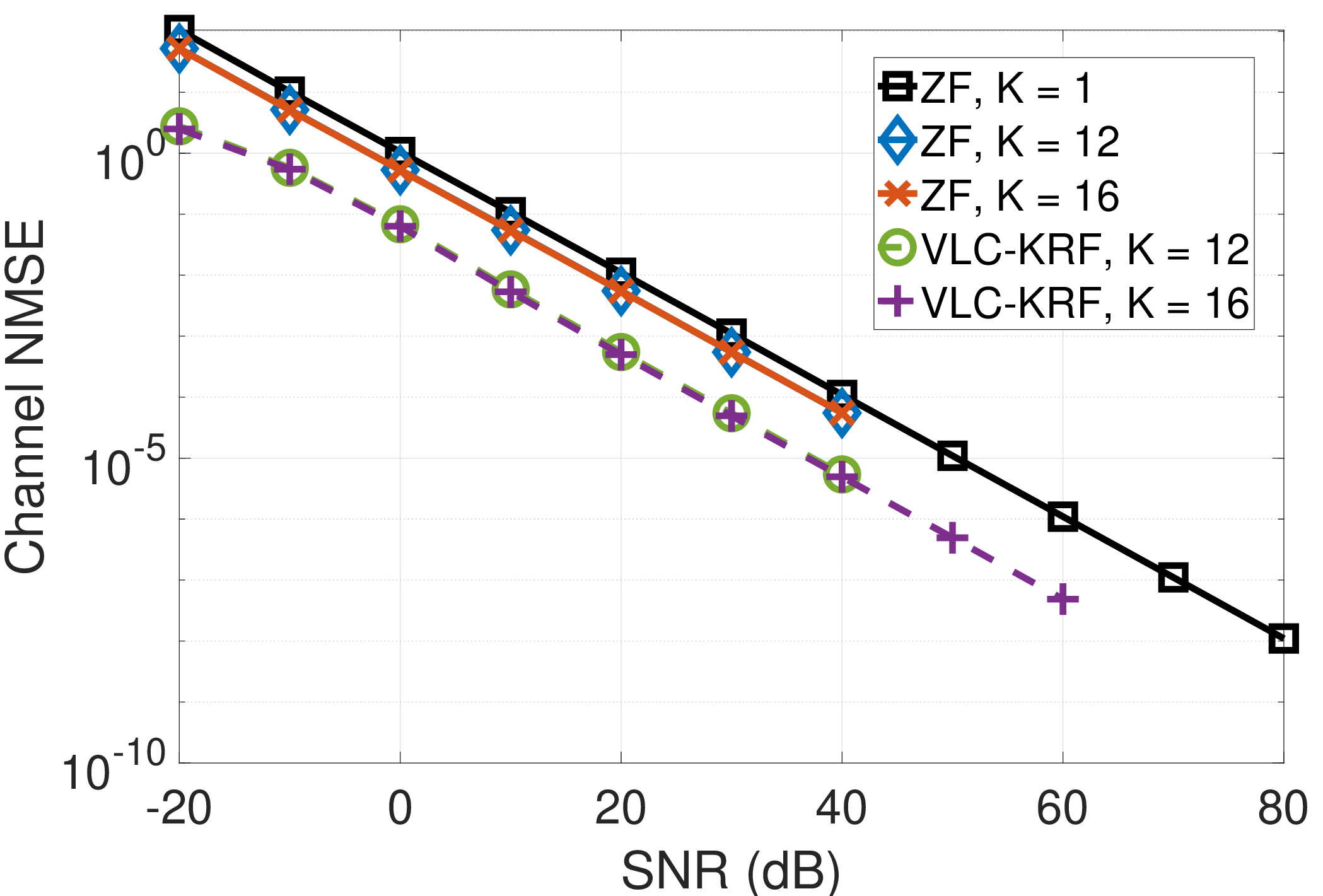}
\label{fig:SNRNMSE}}
\caption{BER and NMSE for the QLED MIMO $2\times2$ system.}
\end{figure}

Fig.~\ref{fig:SNRBER} shows the BER for different values of $K$. Increasing it provides diversity and coding gains, enabling a BER of $10^{-4}$ at approximately $24$~dB, whereas conventional CSK does not reach $10^{-3}$ even at $60$~dB. For $K=12$, the ZF receiver provides an additional gain of nearly $10$~dB over VLC-KRF, at the cost of lower spectral efficiency since it requires $K_TL_T$ pilot time slots for channel estimation. Increasing $K$ also introduces more temporal diversity, but reduces throughput.

\begin{figure}[!t]
	\centering
	\includegraphics[width=0.8\columnwidth]{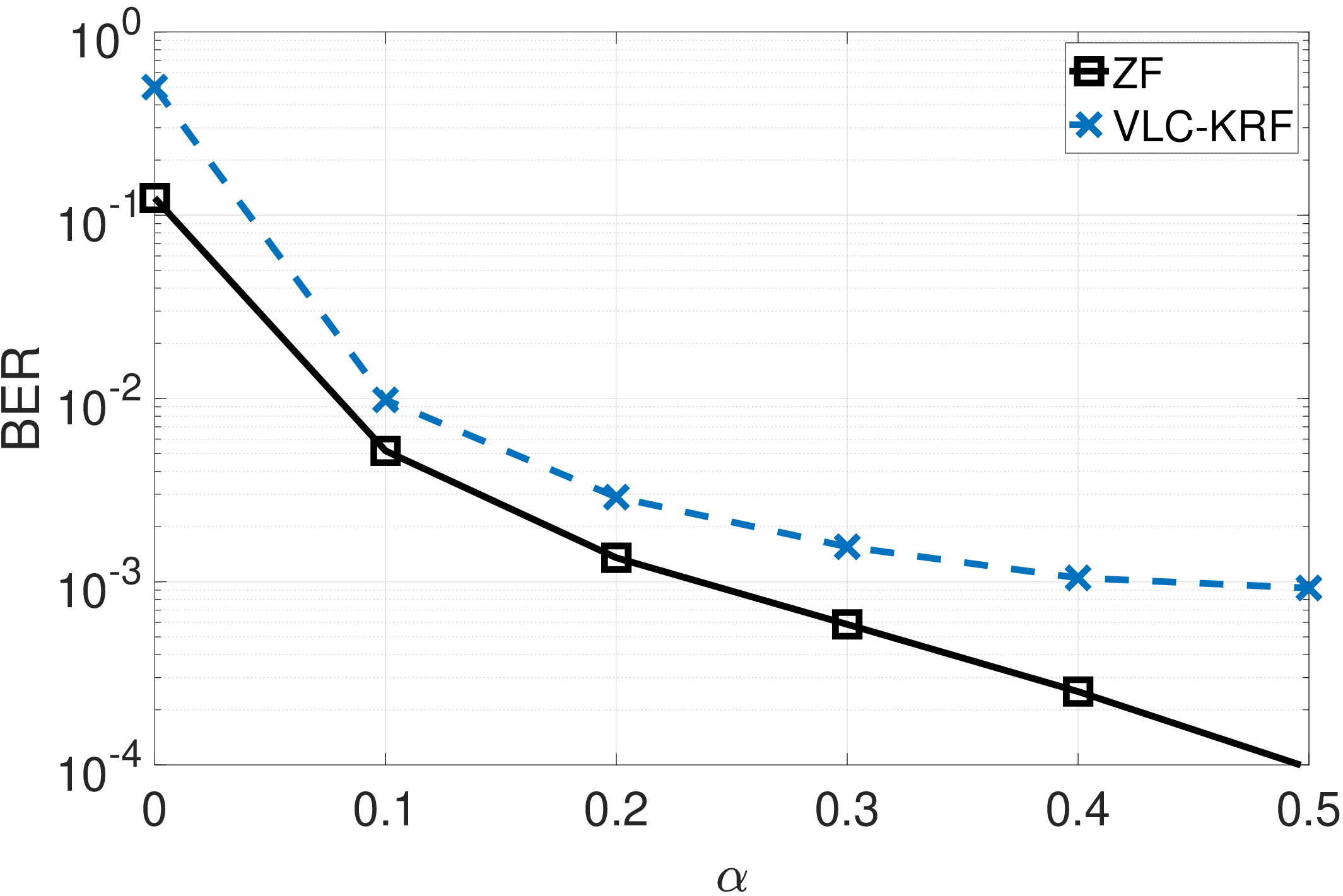}
\caption{BER for different $\alpha$ values in a QLED $2\times2$ system.}
	\label{fig:alphaBER}
\end{figure}

Fig.~\ref{fig:SNRNMSE} shows the normalized mean-square error (NMSE) of channel estimation for both proposed receivers. The VLC-KRF receiver achieves a lower channel-estimation error than the ZF receiver, which is particularly relevant for applications that require accurate channel knowledge, such as visible light positioning. The final simulation evaluates the impact of the parameter $\alpha$, which controls the peak-to-average power ratio of the transmitted signal. A QLED $2\times2$ system with $K=12$, $P_m = 0.5$, and SNR of $20$~dB is considered. The BER performance of both proposed receivers is compared. From a perceptual standpoint, if the modulation rate is sufficiently high - that is, above the flicker-fusion threshold, typically on the order of $100$-$200$~Hz \cite{IEEE2015} - these power variations are not perceptible to the human eye, which integrates the received light and perceives a constant illumination proportional to the average optical power.
As shown in Fig.~\ref{fig:alphaBER}, increasing $\alpha$ improves the performance of both receivers by improving the conditioning of the dimming matrix $\mathbf{C}$ and of the effective channel. The average condition number decreases from $20.9169$ for $\alpha=0.1$ to $9.2635$ for $\alpha=0.5$. The ZF receiver benefits more from this improvement because it explicitly relies on effective channel inversion, whereas VLC-KRF is more robust due to its SVD-based factorization.

Table \ref{tab:eta_comparison} compares the spectral efficiencies of both receivers for different scenarios using the equations \eqref{eq:SE1}-\eqref{eq:SE3}. The gain increases with the number of transmit arrays and decreases as $K$ increases, indicating that the tensor-based VLC-KRF receiver is more advantageous for large-scale MIMO systems, whereas the ZF receiver may offer better performance in smaller MIMO scenarios with a lower spectral-efficiency penalty.

\begin{table}[!t]
\centering
\caption{Spectral efficiency for different configurations ($N=10$)}
\scriptsize
\setlength{\tabcolsep}{6pt}
\begin{tabular}{c c c c c c c}
\hline
Case & $K_TL_T$ & $K$ & $\eta_{\text{ZF}}$ & $\eta_{\text{VLC-KRF}}$ & Gain (\%) & Scenario \\
\hline
1 & 6  & 8  & 0.4651 & 0.4938 & 6.1 & $K_T{=}3,\ L_T{=}2$ \\
2 & 18 & 20 & 0.5505 & 0.5970 & 8.4 & $K_T{=}3,\ L_T{=}6$ \\
3 & 30 & 32 & 0.5714 & 0.6231 & 9.0 & $K_T{=}3,\ L_T{=}10$ \\
4 & 8  & 12 & 0.3125 & 0.3306 & 5.8 & $K_T{=}4,\ L_T{=}2$ \\
5 & 8  & 16 & 0.2381 & 0.2484 & 4.3 & $K_T{=}4,\ L_T{=}2$ \\
\hline
\end{tabular}
\label{tab:eta_comparison}
\end{table}

\section{Conclusion}

This work proposed a DSTC for VLC systems based on CSK modulation, exploiting the multidimensional structure of the problem to enable illumination control while preserving color and satisfying physical constraints. The proposed framework supports two receivers: a ZF receiver with pilot-based estimation and a tensor-based receiver that performs semi-blind joint estimation via PARAFAC. The results demonstrated increased diversity and a significant reduction in BER compared with conventional CSK. In addition, the tensor-based receiver achieved lower channel-estimation error and higher spectral efficiency by reducing training overhead to a single time slot, making it particularly attractive in scenarios with more transmitters. Future work includes investigating codes that jointly exploit color diversity and dimming, as well as exploring alternative tensor models for this scenario.

\renewcommand\baselinestretch{.86}

\bibliography{mybibfile}
\bibliographystyle{IEEEtran}

\end{document}